# From Semiotics of Hypermedia to Physics of Semiosis: A View from System Theory

V.V. Kryssanov and K. Kakusho

**Introduction**

With the recent advents of the World-Wide Web, multimedia, and ubiquitous computing technologies connecting an incredibly huge and still rapidly growing number of heterogeneous information resources, the task of supplying users with an adequate means for accessing these resources – the task of *hypermedia system* design – has become too difficult to solve, relying merely on empirical- and common sense- level decision procedures. A hypermedia system (a relatively new term that usually refers to a user interface of a distributed multimedia database) deals with information organized and stored in network structures consisting of nodes (which are representations of different media types – sound, texts, images, video, etc.), links (which implement physical passages between nodes), and scripts (which direct the time dynamics of links), and it thus enables *hypermedia-based communication* – the cognitive processing by the system's user that includes selecting, organizing, and integrating information represented over time in the connected nodes (Jackson 1997). Since the invention of hypertext – the predecessor and, in a sense, degenerated case of hypermedia, many attempts have been made to formulate a scientific theory that would provide systematic guidelines for the design and development of hypermedia systems (Bush 1945; Nelson 1967; also see Bardini 1997 for an overview).

It may be said that a theory is satisfactory as far as it allows for comprehensive analysis and prediction of certain phenomena or else guarantees a pragmatic application. In the case of hypermedia systems, the phenomena in focus are human communication phenomena. It is then only natural that a large group of design approaches endeavor to apply the Shannon-Weaver Communication Theory by borrowing its conceptual apparatus and analytical procedures to develop information systems for efficient and effective communication (e.g. de Souza 1993). Another large and alternative group of design approaches build on various methodologies (e.g. those of the Object-Oriented Design paradigm), which pay little attention to understanding the processes underlying human communication but, instead, strive to generalize and extend the best-known software engineering practice to new design cases (e.g. Beaudouin-Lafon 2000). While this twofold classification may appear too coarse and simplistic, it should suffice to justify our claim that most of the currently available approaches to hypermedia system design can hardly be considered satisfactory.

By formalizing communication as the transmission of meaning – a perception or idea – from a sender to the receiver, the classical (Shannon and Weaver's) communication theory has made statistical modeling a powerful tool for the analysis of various communication processes. However, as long as there is no brain scanning device that would explicitly reveal what is the meaning sent and received, this theory will fail to provide a reliable analysis of human communication. On the other hand, by substituting a theory with a methodology – no matter how general or 'natural' – one never touches the theory *per se* but is forced to continually adapt methodological procedures to a potentially infinite number of communication situations and develop 'meta-methodologies' (vernacularly called 'ontologies' in computer sciences) to systematize these situations that, in practice, result in unmanageably complicated and contrived design models.

It is likely that any theory of information system design has to be based on a theory of communication. The difficulties with the development of such a theory arise from the apparent necessity to concurrently address two seemingly incompatible aspects of computer-mediated communication: communication as a physical – technological, measurable, and fairly predictable – process, and as a mental – in effect experiential, partially unconscious, and observationally emergent – phenomenon. A communication theory has therefore to locate itself somewhere between two perils of category error: the error of models identifying communication as a mere physical act, and epiphenomenalism – theories struggling to remove any physical grounding (particularly causality) when describing communication. In seeking to avoid the evident dangers, scientists investigating communication and, lately, information and hypermedia systems, have been quick to adopt philosophies, specifically semiotics, that steer clear of the absolutization of either the physical or mental by recognizing and explicitly defining connections between these two, proceeding from the very conceptual level. Two questions, however, come up at this point: whether the mental can be approached in the same way as the physical (principally, can it be in any way 'measured' and predicted) and if so, is a communication theory to be a distinct scientific discipline with its own conceptual and philosophical basis and mathematical tools, but not an empirically tailored unstable compromise between technology, physiology, and the humanities – psychology, sociology, and linguistics? In the present article, we investigate these questions.

Information and hypermedia system design theories are divided according to a transmission/interpretation distinction, and the weak and strong points of each group are examined. Next, we formulate definitions and propositions on a concise but transparent axiomatic basis, which allow us to uniformly describe the mental, physical, and social phenomena of communication. In so doing, we proceed from the assumption that although it may be difficult to unequivocally define communication from positions of a single discipline, it will always involve some autonomous behavior (e.g. human-human and human-computer interactions) constituting a communication language, as well as autonomous dynamic environments – physical and social that impose constraints – principal and potential – on this language. The latter determines our choice of system theory and sociology as the source of conceptual ideas; semiotics comes as a general philosophy to effectively organize all the different concepts within one coherent framework, and physics provides major inspirations as to how the communication mechanism would 'work,' both qualitatively and quantitatively. Keeping the technicalities as few as possible and the notations as simple as possible, we give nonetheless a rigorous mathematical interpretation of the proposed conceptualizations that in fact constitutes the core of what we would call 'Quantitative Semiotics.' We then derive a statistical model that explains and predicts fluctuations of representations in human communication and attempt to not only conceptually but also statistically justify the equivalence of Peircian and de Saussure's semiotic modeling, which initially emerges as merely a 'by-product' of the proposed axiomatization. This article describes two experiments conducted to illustrate and validate the developed approach and discusses the theoretical findings in a context of prior research.

## Communication Models and the Design of Information Systems

Most of the modern approaches to the design of information systems utilize one of the two contending, though not completely alienated conceptualizations of the early Stimulus-Response Model of behavioral psychology, which set about with the communication process as either the transmission of messages or the development of meaning. The modeling of the message transmission goes back to Shannon and Weaver's mathematical theory of communication and its 'conveyor-tube' framework first proposed half a century ago for solving the technical problem of excessive physical noise in communication channels (Shannon and Weaver 1949). In the past decades, this model has significantly been improved upon and adapted to a variety of domains, making it a widely-considered general model for communication studies.

Notwithstanding the changes made to date, the basic concept of communication as transmission of messages remains the core of the approach, and in whatever terms a transmission-focused model is described, whether they be probabilistic, software engineering, or semiotic, it postulates that:

- there is an active source of information – the sender who 'encodes' her or his idea or perception (in other words, meaning) into a message;
- the message is sent through a channel (or medium) to its destination – the receiver who 'decodes' the message and provides feedback;
- upon completing the transmission attempt, the sender's and the receiver's information may differ, owing to the noisy channel-medium and/or alterations caused by the encoding and decoding;
- besides the sender and receiver, there is an observer, which can yet be identical to either of the communicating two, who intervenes in the process by determining the successfulness (or efficiency) of communication through comparison of original and received meanings.

Figure 1 illustrates the concept of communication as the transmission of messages in hypermedia system design. The operation of an information system (e.g. a multimedia database) is locally realized through its interface, which is composed of messages. These messages can be about the system domain, the computational domain, and about the user's possible interactions with the interface. Both the system and the user can send and receive information, and a computational data model implemented in the computer program plays the role of an observer, determining the 'correct' (i.e. successfully 'understood') messages and interactions. This data model reflects an individual's (e.g.

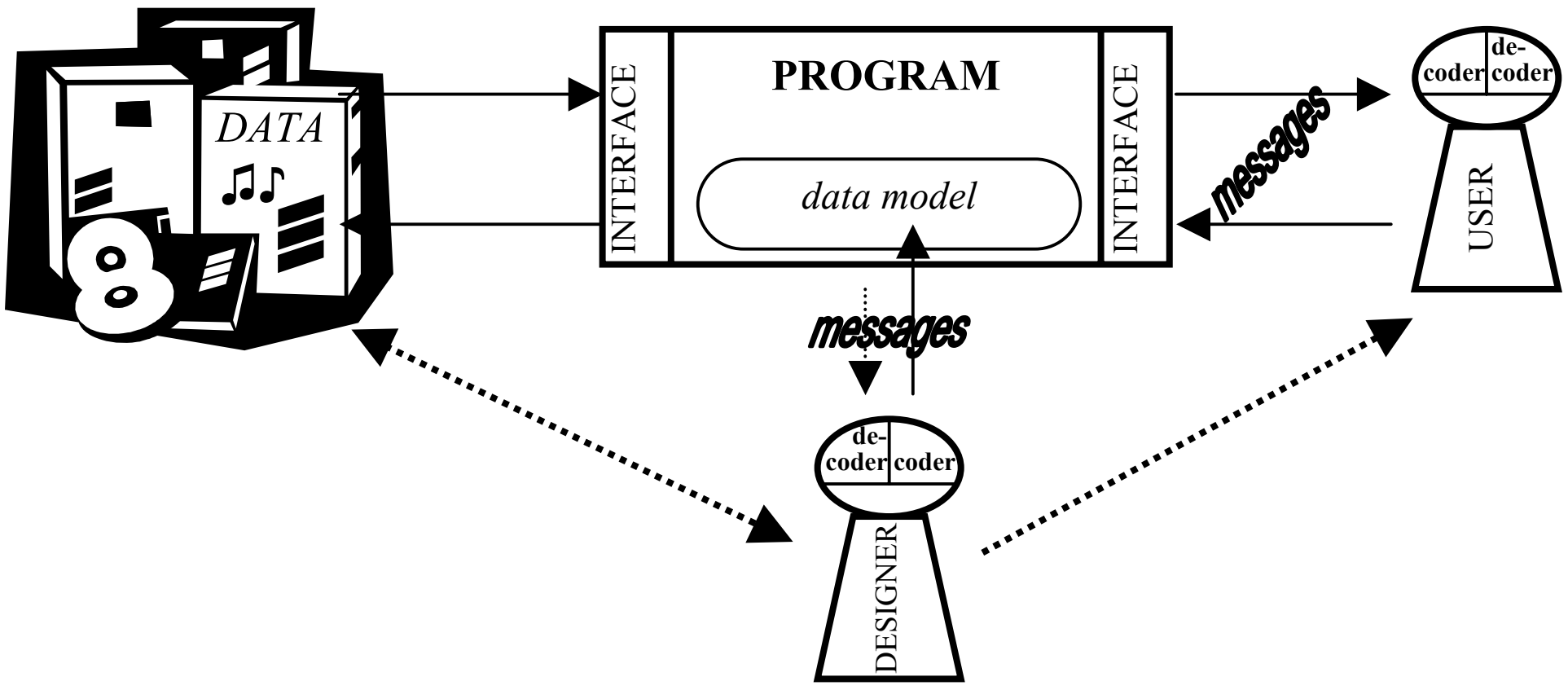

**Figure 1**: Information system and the transmission of messages.

the designer's) comprehension of the domain, structure, and functioning of the system that is encoded and sent to the user through the interface.

In his work (Goguen 1999) on a distributed system user interface, Joseph A. Goguen, a computer theorist, gives a formally well-elaborated and intuitively appealing example of the application of a transmission-focused model of communication. To mathematically define the model, the author develops the novel technique called 'algebraic semiotics,' which puts forward an algebraic interpretation of de Saussure's Semiology (de Saussure 1974). A user interface is characterized in terms of a 'semiotic morphism' that is a mapping (i.e. translation or re-representation) from a 'cognitive' sign system representing the communication situation (e.g. as conceived by the designer) to an 'external' sign system (e.g. a language) representing the user interface, where a sign system is a formal concept somewhat resembling the notion of context-free grammar. It is argued that determining properties of the corresponding semiotic morphism can help evaluate the user interface functionality and quality. In particular, it is argued that in human communication, morphisms preserving structure have a higher priority and are (ethnomethodologically) more important and 'better' than morphisms preserving content. The paper describes the application of the devised model to designing user interfaces of a distributed collaborative system, and it reports the experiences with regard to using them. While the developed interfaces retain logical consistency even in rather complicated operational situations, it remains to be seen why and under which circumstances they would provide efficient and effective communication.

The author sought to justify the design solutions through studies of ethnography, social science, and linguistics which build on characteristics of communication that are average or common for a certain community, e.g. a group of system users. The appeal to semiotics, however, appears weak: the paper does not explain but rather makes it difficult to see if and how anything (but the terminology?) of this discipline can assist the designer in coping with the complexity of communication begotten by numerous socially converging but still experiential, subjective, and never uniform processes of information transmission. Furthermore, by utilizing the approach ‘as is,’ one could hardly rationalize the introduction of the cognitive and social ‘dimensions’ (i.e. the two distinct sign systems) for modeling the communication process.

Communication models of the second, interpretation-focused class concentrate on the process of *signification* – the perception of various objects by humans. With conceptual traces back to the Middle Ages, these models are nowadays habitually associated with the Peircean vision of a semiotic triad connecting a (physical) sign (i.e. representamen) with its object (i.e. that what it signifies – a signified in de Saussure’s terms) by meaning (i.e. signifier), which may or may not be another sign, that is the conceived sense made of the first sign (Peirce 1998). In information system design, the application of such a model (frequently simply called ‘semiotic model’) is concerned with the generating and exchange of meaning, when the divergence of (sent and received) meaning is not a failure but a natural attribute of communication. In a semiotic model, communication is dealt with as the development and re-interpretation of signs that are

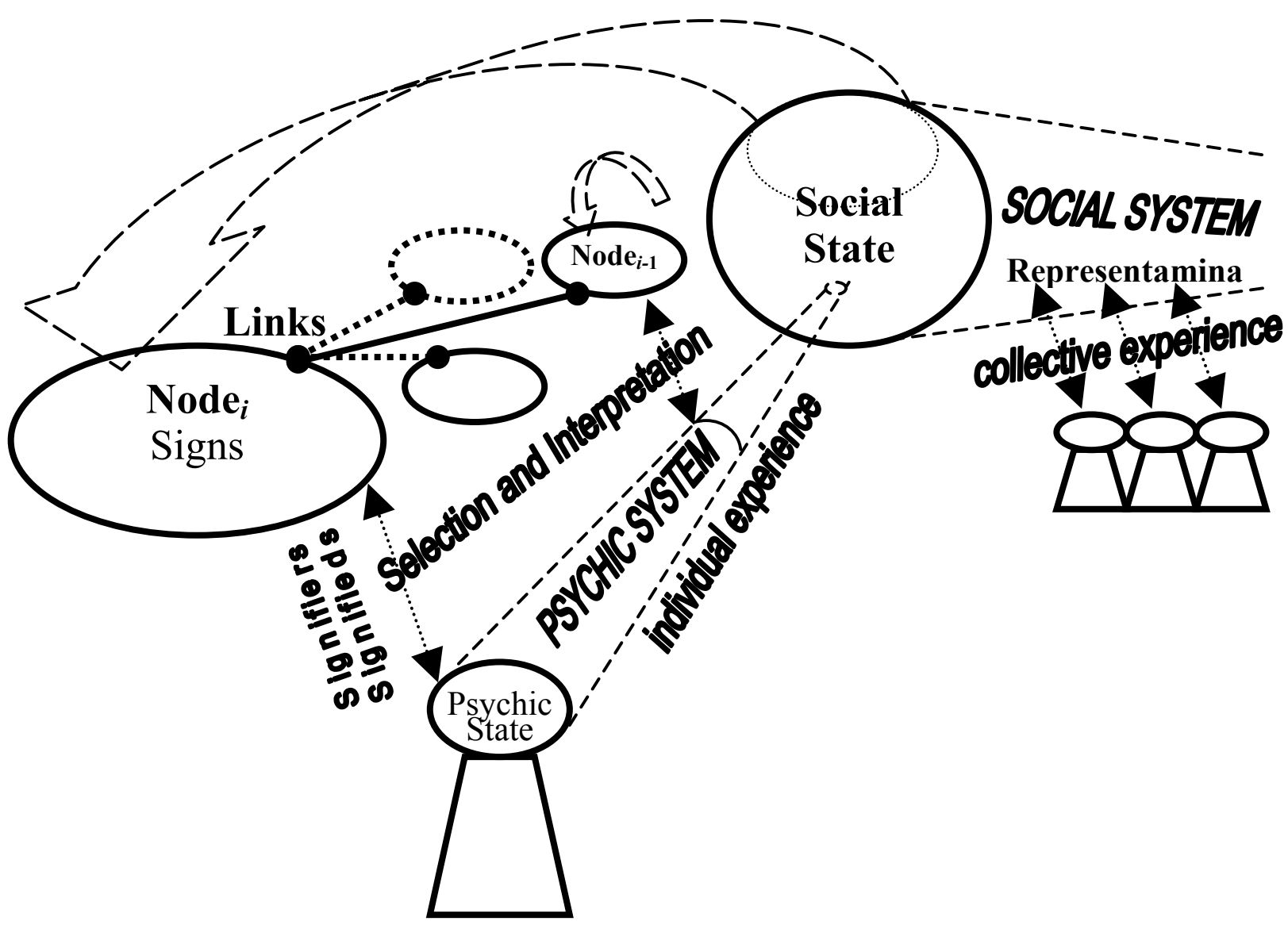

**Figure 2:** Semiosis of hypermedia-based communication.

representations of physical or mental entities. A semiotic model of communication postulates that:

- there are no senders and receivers – they are uniformly represented and effectively replaced by interpretants, i.e. by those which follow semantically from the processes of interpretation;
- not a message but meaning emerges, which is externalized and (re)determined through interactions between a carrier (e.g. gesture, text, picture, sound, and the like) and the culture (i.e. the social system standing for it);
- decisive notions, like ‘correct’ or ‘successful,’ are subject to both individual (i.e. experiential) assessment and socio-cultural convergence or compromise.

Figure 2 depicts a semiotic model of hypermedia-based communication. The interface is built of nodes comprised of signs. The signs are to represent the states or, in other words, ‘meaning’ of psychic (or *conscious*) systems. This meaning is externalized with signifieds (generally seen as behavior; in other words – objects), which may prompt the ‘generation’ of signifiers (i.e. signs standing for something, or interpretants) through interaction with the social system. The interface is a (partial) realization of the social system state, while signs are individually interpreted but necessarily have socially- and culturally-induced meaning. There can be more than one social system but, tautologically, no communication is possible beyond a social system. All the systems involved have internal (time-)dynamics affecting their output: signifieds/objects in the case of psychic systems, and signifiers and signs in the case of social systems.

Peter B. Andersen sketched the range and applicability of the semiotic models for human-computer interaction (Andersen 1990). It was argued that whenever computation and interpretation have an effect on each other, they can and should be studied semiotically, and that not only straight human-computer interaction, but also any storage and retrieval of data are communicative processes. Building on the knowledge of older and traditional semiotic disciplines, such as literature and art, the author discusses and elaborates on the following points: *i*) the interface of an information system should agree with and expose what really goes on behind the interface, i.e. it should reveal the 'actual' semantics of the reality and hence be grounded upon the experience of sense; and *ii*) apart from being interpretable (at least potentially), the interface should be verbalizable in terms of the user's work-language that may be seen as a realization of the corresponding social system.

The recent work (Andersen 2001) provides a design example that illustrates these two issues. It is shown that projecting features from verbal communication onto the user interface can enhance the communicability of a multimedia computer system. The developed interface is expected to be effective and efficient for a social group delineated by the users' ability to comprehend the language employed. A serious drawback of this and many other semiotic (i.e. interpretation-inclined) models is, however, that specific concepts, the terms with which the models are defined, are based on introspection and therefore lack formal precision that makes it problematic to use them on a computational basis (e.g. see Schmidt-Isler 2000; Condon 2000). This, to a large extent, results from the noticeably confusing array of contemporary semiotic theories, which build on diverse, personal, and formally vague but intricate conceptualizations while describing the same phenomena.

Being quite aware of the risk of perplexing the reader with the terminology, we nevertheless deliberately presented two examples, where the conceptually dissimilar but semiotic terms were deployed to formulate the principles for designing hypermedia systems. Our motivation is that most naturally, communication is indeed a semiotic phenomenon, and the two perspectives – one focused on the transmission, and another on interpretation – correspond to two distinct but overlapping parts of semiotics named by Umberto Eco as Theory of Sign Production and Theory of Codes, respectively (Eco 1976). Semiotics offers a common basis for discussing, analyzing, and designing information systems, and the two different approaches to modeling communication can be related to studying two aspects of interpretation of the same sign: as an object which is interpreted (e.g. a sign in Goguen's sign system, which is, in itself, a system of interpretance) and as an object of interpretation (e.g. meaning in Andersen's terms). The latter could also be posed as an ordered pair of objects interpreted, which are defined on two systems of interpretance accommodating syntax for one element of the pair, and semantics for the other (e.g. as in Sonesson 2002).

Now to summarize: each of the popular design approaches surveyed in this section has serious limitations in modeling human communication. Just as the use of the semiotic terminology alone cannot make a statistically justified conveyer-tube framework capable of handling the complexity and subjectivism of hypermedia-based communication, even a thorough semiotic analysis cannot 'automatically' make the task of hypermedia system design more understandable or more easily formalizable and thus controllable. The separate treatments of the different aspects of interpretation eventually led the authors of these, as well as many other modeling approaches, to the study of information (re)representation with little attention to the measurable and reproducible (i.e. epistemologically objective, as opposed to introspective) characteristics and consequences of communication that, in many practical cases, made analyzing and designing hypermedia systems *ad hoc*, ambiguous, and largely unpredictable.

**('Quantitative') Semiotics of (Hypermedia) Systems**

The strongest point of the Stimulus-Response Model and its classical (Shannon and Weaver's) mathematical interpretation is that it allows for statistically modeling and therefore predicting the effect of a single act of message transmission (or series of such acts unchanging in condition), provided that one can measure 'the amount of' meaning, the efficiency of its coding, or determine the 'noise properties' (i.e. to what extent it would alter the conveyed perception or idea) of the medium-channel. Looking at the process more carefully, however, one quickly finds that this model does not and cannot work in a general case, e.g. in hypermedia-based communication, where the receiver's meaning appears a result of the cognitive processing of a number of stochastically independently generated and often incomplete messages, rather than one stationary decoding process. The principal problem is that the Stimulus-Response Model cannot, in its canonic form, account for the fact that meaning is not transmitted from a single source, but created in the mind of the receiver (i.e. interpreter), based at best in part on the sender's message but also – on social and idiosyncratic parameters of the communication situation, and yet sometimes regardless of the sender's original intention. In seeking to overcome this modeling deficiency, we will first re-define communication in such a way as to avoid the necessity of conjecturing about and (directly or otherwise) assessing the ‘inner contents’ (i.e. possessed information, meaning, etc.) of the communicating parties to estimate the efficiency (correctness, etc.) of the process of meaning ‘exchange’.

From positions of evolutionary structuralism (see Maturana and Varela 1980), communicative activities can be grouped into a specific category of observed behavior of self-organizing systems, such as humans or other animals. The principal property of a self-organizing system is its autonomy in respect to the environment: the system state at any time is determined solely by the system’s structure and previous state. The system cannot be controlled from the outside, and environmental perturbations can only be a potential cause for the changing of the system state. Hence, all observed behavior – the output – of a self-organizing system is a result of its inner state and history. Through behavior, the system can interact with the environment that may cause it to change its structure, so that the system becomes structurally coupled with the environment. It is said that the coupled system undergoes self-adaptation, when the system and its dynamic environment mutually trigger their inner states. The self-adaptation processes of several systems embedded in the same environment may become coupled, recursively acting through their own states. All the possible changes of states of such systems, which do not terminate this coupling, establish a consensual domain. Behavior in a consensual domain is mutually orienting. Communication can fundamentally be defined as the *observed behavioral coordination* developed from the interactions between autonomous self-organizing systems *in the consensual domain* (di Paolo 1998).

Inspired by the conceptual compatibility of system theory, quantum physics, and semiotics (see Nadin 1999 for a relevant discussion), we will begin formalizing the system-theoretic account of the communication process by introducing an axiomatic basis as follows.

**<u>Axiom I.</u>** Each autonomous system can be represented by $\Xi$ its inner state space. The (inner) state of the system is completely described by $\gamma$ an attractor basin – a subset in $\Xi$. ■

In this paper, the inner state space is understood as equivalent to the phase space – an imaginary, not necessarily completely determined but representable map of all the possibilities open to the system.

**<u>Definition 1.</u>** Two states of a system, $\gamma_1$ and $\gamma_2$, are called orthogonal, written $\gamma_1 \perp \gamma_2$, if $\gamma_1$ implies the negation of $\gamma_2$, or vice versa. ■

**<u>Definition 2.</u>** For a subset of states $\Gamma \subset \Xi$, its orthogonal complement is

$\Gamma^{\perp} = \{\gamma \in \Gamma \mid \forall \gamma' \in \Gamma^{\perp} : \gamma \perp \gamma'\}$. ■

**<u>Definition 3.</u>** $\Gamma \subset \Xi$ is orthogonally closed if $\Gamma = \Gamma^{\perp\perp}$. ■

Accordingly, *the orthogonal closure implies the existence of a distinction*; non-orthogonal states can only be distinguished in a statistical sense.

**<u>Axiom II.</u>** Each observable state (i.e. representation or behavior) of the system can be specified as $\alpha$ an orthogonally closed subset of the attractor basin $\gamma$, $\alpha \subset \gamma$, $\alpha = \alpha^{\perp\perp}$, and as the smallest such set. ■

Somewhat similarly with a quantum system, the autonomous system is *observationally in multiple states simultaneously*: at every single moment, more than just one representation of the system state can be made. This is formally elucidated with the following theorem:

**<u>Context Theorem.</u>** For every two distinct representations $\alpha \neq \beta$, $\alpha \subset \Xi$ and $\beta \subset \Xi$, there exists a context representation $\delta \subseteq \alpha \cup \beta \subset \Xi$ such that $\forall \omega \subset \Xi$, if $\omega \perp \alpha$ and $\omega \perp \beta$, then $\omega \perp \delta$.

**Proof:**

Let us consider two states $\gamma_1 \neq \gamma_2$ characterized by two representations $\alpha \neq \beta$, respectively. One can virtually always define an observable $\pi \subseteq \alpha \cup \beta$ characterizing some $\gamma_3$, $\gamma_3 \neq \gamma_1$ and $\gamma_3 \neq \gamma_2$. Since $\pi \cap \alpha \neq \varnothing$ and $\pi \cap \beta \neq \varnothing$, $\pi$ cannot be determined orthogonal to $\alpha$ or to $\beta$. Furthermore, for all the states $\gamma_4$, $\gamma_4 \perp \gamma_1$ and $\gamma_4 \perp \gamma_2$, with representations $\omega$ different from both $\alpha$ and $\beta$, $\omega \cap (\alpha \cup \beta) = \varnothing$, $\gamma_3 \perp \gamma_4$ as $\gamma_3 \cap \gamma_4 = \varnothing$. Hence, $\pi$ is a *context representation* of $\alpha$ and $\beta$. ■

**<u>Definition IV.</u>** The dynamics of coupled autonomous systems are given by an *n*-tuple of pairs of recurrent equations defined as follows:

$$\begin{cases} R_{k,j+1} = \mathbf{E}_k(S_{k,j}, t), \\ S_{k,j+1} = \mathbf{I}_k(R_{k-1+n\delta_{in},j+1}, t), \end{cases} \tag{1}$$

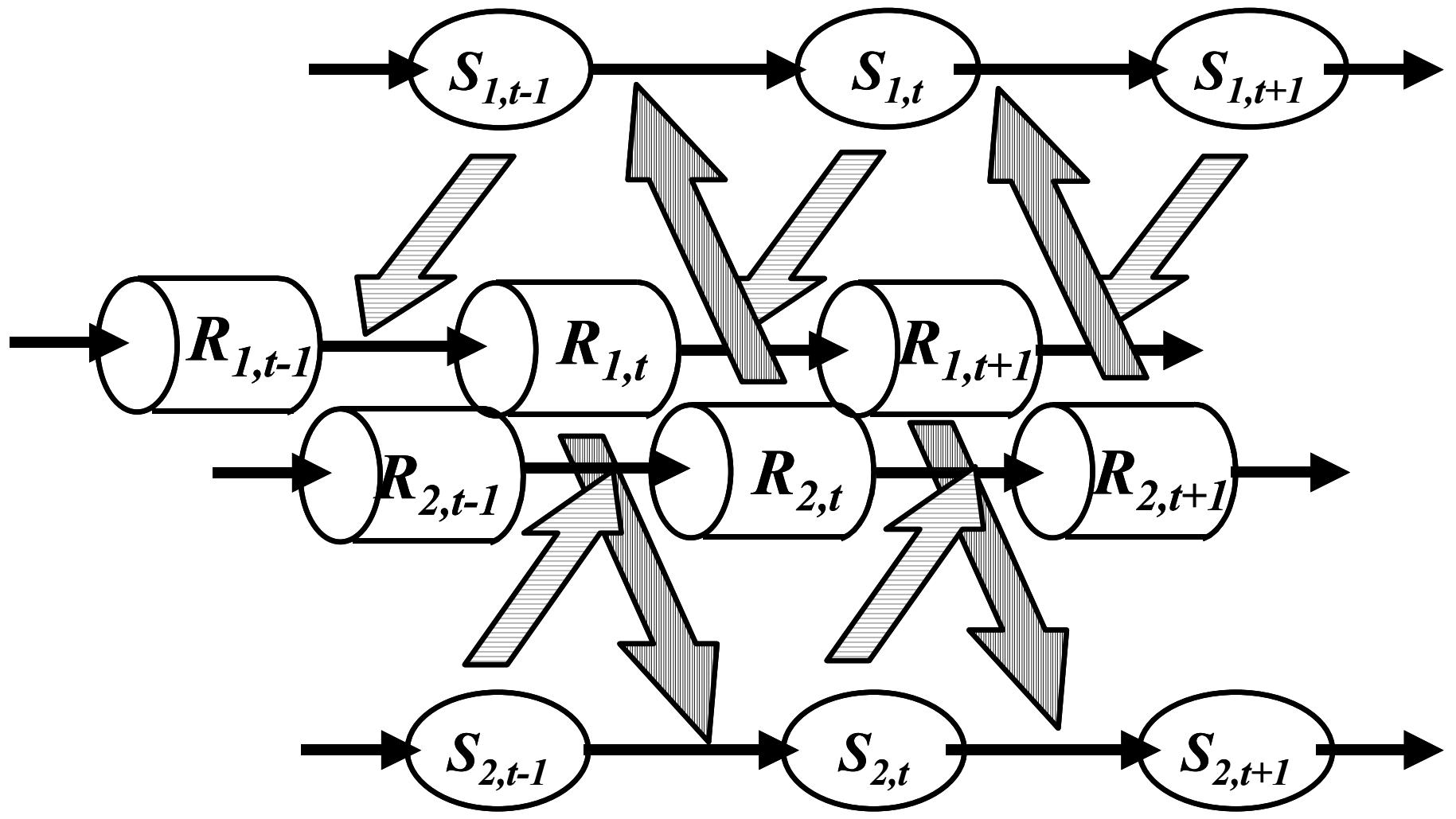

**Figure 3:** Coupling of autonomous systems (the case of $n$ = 2).

where $R_{k,j}$ denotes a representation (i.e. observable or *observed realization*) of $S$ an inner state of a $k$-type system at $[t_{j-1}, t_j]$ a discrete time interval, $t_{j-1} < t_j$, $j$=1,2,…, $k=n-i+1$, $i=1,...,n$; $n$ determines the *depth of coupling*, $\delta_{in}$ is the Kronecker delta: $\delta_{in}=1$ iff $i=n$, and $\delta_{in}=0$, otherwise; **E** and **I** are time-dependent operators specifying dynamics of the coupled systems at the macro- (as for an observer) and micro- levels, respectively. ■

Equations (1) provide a formalization of the system-theoretic portrayal of communication as mutually-orienting behavior of coupled autonomous systems (also see Figure 3). To move on with the modeling and maintain the discussion at a reasonably straightforward and conceivable level of abstraction, it is now convenient for us to give a semiotic interpretation to the set-theoretic concepts introduced above. For the sake of simplicity (but with no loss in generality), we will employ Peirce's triadic scheme for a sign which is then the relationship between a sign vehicle (i.e. the physical sign), its object (i.e. the signified entity), and its interpretant (i.e. the result of an interpretation of the physical sign).

**Definition V.** In terms of Definition IV, we will call *object* an observable of an inner state of a $l$-type autonomous coupled system. The inner state of an $p$-type system, $p \geq l$, engaged in the self-adaptation process is an *interpretant* that may be related with the object through its (the object's) *sign* that is an observable of an inner state of an $m$-type system, $m \geq l$, $m \neq p$; $l, p, m = 1, 2, \ldots$ . ■

Definition V dictates *a necessary condition* for semiotic triads: there must be at least two systems, each of a different type, to ensure the existence of the object-sign-interpretant relationship. Obviously, *a sufficient condition* is the existence of a relation (e.g. associative) between the sign and the interpretant. The latter can also be understood as a synchronicity constraint for the time-intervals: the sign and its interpretant, though not necessarily its object, must develop simultaneously (this accounts for the fact that, in terms of time intervals $j$, communication is experiential in principle and thus retroactive, not prospective, in respect to the phenomena it describes).

A typical (and the classical – the conveyor-tube) case would be to consider the coupling of autonomous systems of two different types (i.e. $n = 2$, in terms of Definition IV): psychic (e.g. humans) and social (e.g. a language). Observables $R_2$ of the social system – signs – correspond to a socially recognized and *anticipatedly* effective representation (i.e. behavioral pattern) of a concept from $S_2$ a class of possible concepts, while observables $R_1$ of the psychic systems stand for a representation (as 'externalized,' e.g. utterance) of $S_1$ conceived perceptions and ideas. The role of the social system is to filter, or authorize, communication out of human behavior but, on the other hand, to buffer the behavior against the rational uniformity of socio-cultural norms. Owing to the autonomy of the systems involved, the social system does not (and cannot) impose a 'standard' of communicative behavior, but rather serves to propagate among the psychic systems regularities enabling coordination of their behavior. An extended (e.g. Peircian) vision of communication assumes a grounding in the physical world and therefore requires taking into consideration the third type of coupled systems – physical. Signs may, in this case, be seen as mediators allowing some sort of vocabulary learning or acquisition, or self-adaptation induced by the 'laws' of nature (i.e. by the inner states of the physical environment as they are 'externalized' and/or perceived).

At this point, the task of the modeling of communication can effectively be reduced to the task of the specification of semiotic triads, for instance – as they are manifested with signs. Indeed, the sufficient condition for the development of triads stipulates that there is a perceptible connection – coupling – between systems of two different types. Given the definition of the systems' dynamics with equations (1), this coupling will, eventually, result in behavioral coordination of the systems of at least one type, as for any kind of social (i.e. cooperative) activity, these systems will experientially have to 'classify' their shared environment (i.e. the systems of the other type) into a set of attractor basins – recurring behavior clusters or behavioral patterns. In the sense of Definitions IV-V, communication thus assumes the development (though not the *a priori* existence) of triadic sign relationships.

It is important to note that since at least one element of the triad – the interpretant – is, as stated by Axiom II, observable always only to an extent, any communication has to be representationally uncertain in respect to the phenomena it describes (e.g. the inner states of the systems in focus). More formally, this fact is illustrated with the following proposition formulated in semiotic terms.

**Closure Lemma.** A communication is orthogonally closed:

*a)* *pragmatically* through the laws of nature in the sense that given an interpretant $\gamma$, it is only the physical laws determining the observation process that affect the 'choice' of (or association with) its object $\alpha$, so that $\alpha = \alpha^{\perp\perp}$;

*b)* *semantically* through the psychic system in the sense that given a psychic system and a distinction classification of its states by interpretants, it is only this classification – the semantics of objects – that determines the objects (i.e. categorizes or else quantizes the environment) for the psychic system; and

*c)* *syntactically* through the social system in the sense that given a social system, it is only the behavior (i.e. observables) of other type systems – psychic and/or physical – that determines a distinction classification of its states and hence determines the possible signs. ■

In view of Definitions IV and V, the proof of this lemma is a direct corollary of Context Theorem and is self-evident.

The above lemma states that while the inner states corresponding to every physically possible object should, in principle, uniquely determine the communication situation, the object corresponding to each psychic state does not have to be unique. It also entails that in reality, every single communication is orthogonally closed only to a degree. Indeed, given a communication situation, its pragmatic closure can be established if one considers typically a huge number of objects – in fact, all possible objects, which are to express the physical frames of the situation and to precisely grasp the corresponding inner (e.g. physical or psychic) state. The latter is not a practical case unless one considers learning by trial and error or a similar process, and objects are the result of some relations (not necessarily conventions) developed from individual experience – the classifying of the environment by interpretants, rather than exhaustive representations of the related inner states. Moreover, semantic closure is hardly reachable, because to hold, it requires the definition of all the objects for all the inner states. This is unrealistic owing to the spatio-temporal dynamics uniquely allocating each system every time, the indirect (i.e. incomplete, in the sense of Axiom II) character of state assessment, and the natural cognitive limitations (e.g. the memory limits). This, as well as the fact that social systems are generally dynamic in respect to their constituents (i.e. the coupled psychic systems), makes the achievement of syntactic closure highly unfeasible, too. Therefore, *every single communication is representationally uncertain*.

It can be shown that having defined the orthogonal syntactic (or semantic) closure, one can always reconstruct the state(s) of the social (or psychic) system, though not the communication situation formed by the inner states, as there will always be an uncertainty caused by the 'externality' – in respect to the social (and psychic) system – of the pragmatic closure. This once again explicates the principal fallacy of the notion of 'efficiency' in transmission-focused communication models: any estimate based on the comparison of 'sent' and 'received' representations can have an arbitrary meaning (e.g. the 'history' or even 'prediction' rather than description of phenomena), depending on the inner, partially hidden state of the described system (e.g. physical) at the times of the sending and receiving. A measure free of the contingent amendments caused by pragmatic uncertainty could be an estimate of the behavioral coordination of coupled systems.

**Definition VI.** For communicating systems of the same type (as it is stipulated by Definition IV), the efficiency of communication is determined by $COR(t)$ a characteristic of their observed behavioral coordination specified as follows:

$$COR(t) \cong \sum_{t' \neq t} \frac{\mathrm{N}^S(R_t \cap R_{t'})}{\mathrm{N}^S(R_t) + \mathrm{N}^S(R_{t'}) - \mathrm{N}^S(R_t \cap R_{t'})}, \tag{2}$$

where $\mathrm{N}^S(R_t)$ is the number of $S$ distinct inner states of the systems, which have the same observable $R_t$, $\mathrm{N}^S(R_{t'})$ is the number of distinct states having the observable $R_{t'}$, $\mathrm{N}^S(R_t \cap R_{t'})$ is the number of states with both of the observables; $t$ and $t'$ are marks of discrete time intervals. ■

*COR*(*t*) shows how the corresponding closure is changed through communication: as behavioral coordination produces contextual relations between representations (i.e. categorizes or quantizes the environment), it should reduce the number of distinct inner states (i.e. interpretants) associated with the given communication situation or, in other words, it should reduce the uncertainty of communication – increase *COR* – by propagating a particular behavioral pattern resultant of the classifying of the environment (in other words, of the 'narrowing of choice').

*Experiment*

To illustrate the proposed measure of communication efficiency, we have conducted an experiment. The behavioral coordination has been estimated by calculating *COR*(*t*) for hypermedia-based communication by an experienced system user (the expert), by a user with no *a-priori* knowledge of the same system (the non-expert), and by the non-expert using an advanced interface of the same system, which is capable of adaptation to the user's language, based on feedback. The words comprising the interface – 'keywords' – were processed as signs (i.e. observables $R$ in formula (2)) by assuming that the documents indexed with them correspond to the inner states $S$ (seen as, for instance, goal-states; one state per every distinct document). Figure 4 shows the results of the experiment for the first 5 signs: the thin marked lines show the behavioral coordination for a typical single communication by the three users, and the bold show the average of over 20 communications on different (but related) topics (some more details on the adaptive interface and the experiment methodology can be found in Kryssanov *et al.* 2002). It should be understood that while *COR(t)* is discrete by its very nature, the lines were drawn to display the characteristic trends.

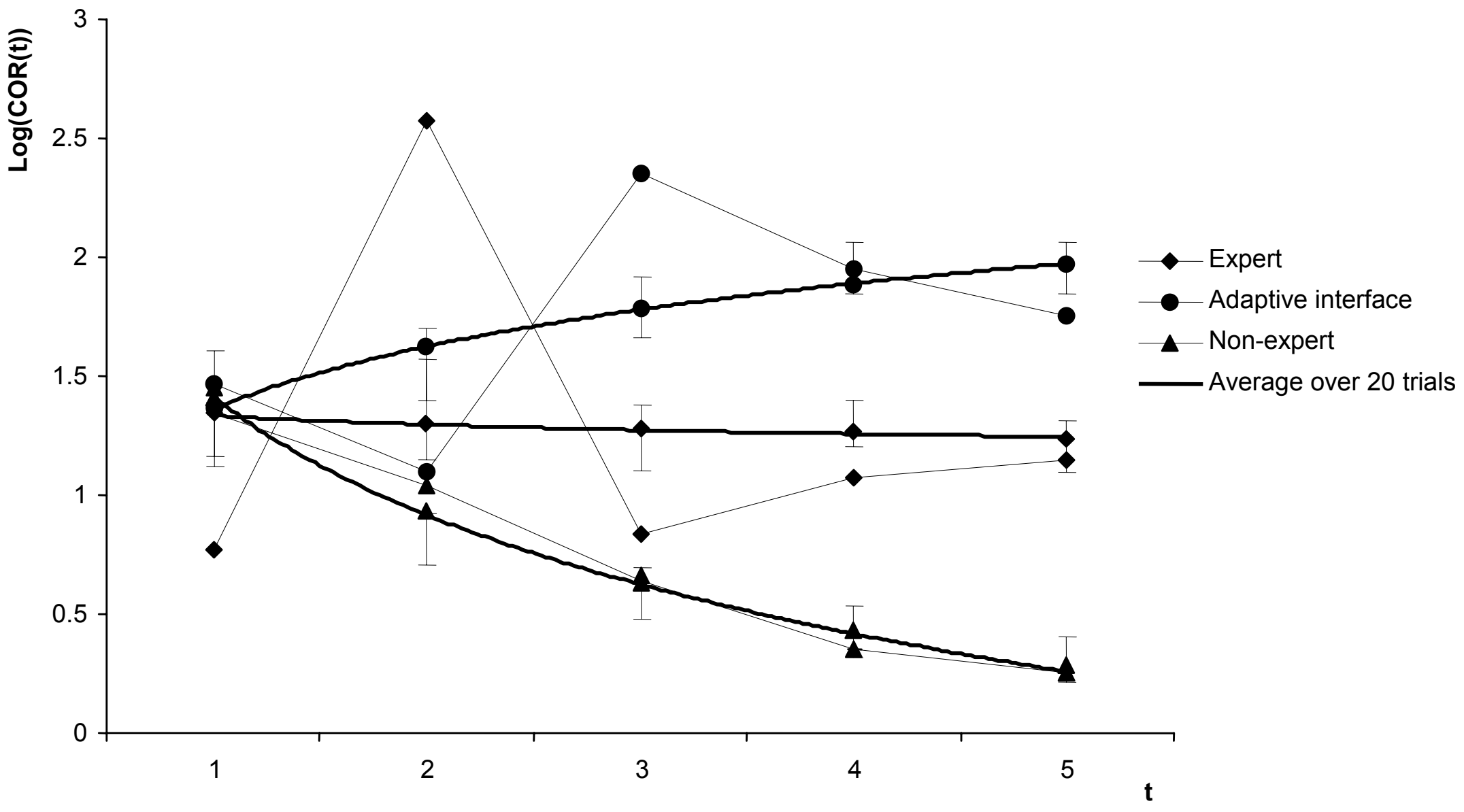

**Figure 4:** Hypermedia-based communication as behavioral coordination.

The figure indicates that the uncertainty of communication by the expert quickly converges to a specific value as in this case, there is no or almost no learning of the (system's) language (and, probably, of the communication topic – the system domain), and this value is likely to reflect the pragmatic uncertainty specific to the given hypermedia system. In the case of the non-expert using the same system, the uncertainty of communication is increased through observation over the examined time interval that may mostly be due to learning about the social language utilized by the system. Finally, the uncertainty of communication by the non-expert with the adaptive interface quickly converges to low values (i.e. to high values of *COR*(*t*)) corresponding to better, even if comparing with the case of the expert, behavioral coordination. For the latter, the question remains if the interface was 'too adaptive' and filtered out too many relevant documents not covered by the user's feedback. This, however, does not refer to the communication process, but rather to the classifying ability of the user and/or the interface. It is evident that a 'good' (i.e. efficient, in the sense of mutual orientation) interface of an information system should provide for low variation of $COR(t)$ (that is convenient to examine using, due to Weber-Fechner Law, a logarithmic scale).

**Towards a Physics of Meaning-Making**

One may already notice that the formalization of the semiotic terms proposed in the previous section with Definitions IV and V implicitly ascertains that Peirce's and de Saussure's conceptualizations of the sign relationship can be considered practically equivalent, unless the modeled communication process (i.e. the coordinated behavior) persistently refers to or is strongly affected by, or tangled with a third-type system, for instance – the physical environment. Reformulating this, the equivalence of the triadic (Peircian) and dyadic (de Saussure's) systems of semiotic coordinates depends on whether the asynchronicity in object and sign (both as observables) dynamics can effectively be neglected, i.e. it depends on the observable (discernable – as detected or measured) depth of coupling. Since no model can be validated theoretically, in this section we will endeavor to find a way for empirical estimation of the efficient (from the standpoint of the modeling simplicity) depth of coupling (i.e. the number of semiotic dimensions) necessary to unequivocally describe communication by investigating the dynamics of observed realizations (i.e. representations) of coupled system inner states. To do so, we have first to elaborate on a mathematical model of the representation dynamics that would also elucidate the conceptual definitions from the previous section.

Let us briefly recapitulate some of our main premises. We postulated that the same (i.e. uniquely characterized by microparameters) inner state of an autonomous coupled system can have different observed realizations (such as, the same meaning can be expressed in many ways, the same sign can signify multiple concepts, and the same object serves as an indicator of various 'internal' and thus hidden processes). In order for communicating systems to guarantee their behavioral coordination can be achieved in a finite time (or, viewing this differently, to guarantee that communication conveys *in principle* finite meaning), the number of observed realizations must be finite for at least one type of the coupled autonomous systems. This makes the communication process observationally cyclic and representationally constrained: recurrence of identical representations is inevitable, given a sufficient observation time (it can well be illustrated by the fact that 'traditional' communication media, such as languages, encompass always a finite number of socially recognized representations – signs).

We will seek to determine the distribution of $z(t)$ the occurrence number of a representation across increasing expenditures of time (e.g. it can be understood as the distribution of the occurrence frequency of a word in a document) that in practice means to determine $\mathrm{F}(z)$ a cumulative distribution function (CDF) and/or its differential analog – $f(z)$ a probability density function (PDF) of $z$ (these functions completely characterize statistical properties of a stochastic variable and are related as $\mathrm{F}(x) = \mathrm{P}(X \le x) = \int_0^x f(t)dt$; P(A) denotes the probability of event A occurring). Under the above assumptions, $z(t)$ can be defined as follows (the reader unfamiliar with or uncommitted to formal statistical derivation techniques can skip ahead to the final result – formula (9) and the description of an experiment at the section's end; the interested reader can otherwise consult one of the standard textbooks on statistical analysis, e.g. Walpole and Myers 1993):

$$z(t) = \theta \tau_0(t), \tag{3}$$

where $\tau_0(t)$ is the *representation rate*, and $\theta$ is the observation time – a behavioral microparameter that we will interpret as interaction tempo.

Owing to Context Theorem, there are multiple representations of a given (inner) state. A number of statistically independent factors, including those due to semantics (i.e. idiosyncratic experience), syntax (i.e. socio-cultural norms), and pragmatics (i.e. circumstantial anticipations and/or effects), determine their rates $\tau_i(t)$, and some of these factors influence the observed realization by increasing or decreasing its rate $\tau_0(t)$. Temporal changes of $\tau_i(t)$ are controlled by a 'competition' process, which can be thought of as *representational decision-making* (conscious or otherwise), and in which different representations compete for the time available to 'connote' or 'externalize' the inner state. Changes of the available time – *representation time dissipation* – can be estimated by calculating the difference in the state and its representation rates. The dynamics of $\tau_i(t)$ can then be approximated with a diffusion process represented by noisy differential equations (the book by Ghez (2001) provides a thorough introduction to diffusion processes; similar diffusion processes have been studied in cellular biology, Czirok *et al.* 1998):

$$\frac{d\tau_i}{dt} = a_i\left(\rho\mu - \sum_{j=0}^{N}\tau_j\right) + \eta_i, \tag{4}$$

where $\mu$ is the inner state rate, $N$ is the number of competing (i.e. potentially available) representations, $i$=0,1,…,$N$; $\eta_i(t)$ is due to a noise-induced (i.e. random) variation in representation rates and is a Gaussian stochastic variable with zero mean; $a_i(t)>0$, and $\rho \geq 1$ is a parallelism (or, looking at it from the other side, redundancy) coefficient – a (medium-driven) microparameter to account for the apparent concurrency in representing the inner state.

The system of differential equations (4) describes the process of meaning (i.e. inner state) *observational* diffusion in the vicinity of $\rho\mu = \sum_{j=0}^{N} \tau_j$ the (imaginary) hyperplane formed by the meaning's possible representations. One can show that, given a sufficient time and $N >> 1$, this process yields an exponential probability distribution for $z$ with $\mathrm{F}_1(z)$ a cumulative distribution function estimated as follows (Czirok *et al.* 1998):

$$\mathrm{F}_1(z) \propto 1 - \mathrm{e}^{-\lambda z}, \tag{5}$$

where

$$\lambda = \frac{\varepsilon}{\theta\rho}, \tag{6}$$

and $\varepsilon = \frac{1}{\tau}$ is the average representation time for $\tau = \frac{\mu}{N+1}$ the average representation rate of a given inner state. (The PDF of an exponential distribution has the following form: $f(z) = \lambda \mathrm{e}^{-\lambda z}$.)

It is remarkable, that despite the oversimplified character of the specification of the 'diffusion' of an inner state over its observed realizations with equations (4), the exponential behavior (5) will hold for a wide class of models, e.g. when some of the parameters assumed herein as 'stationary' or constant commence changing with time, when some of the apparently independent factors determining representation rates prove mutual dependence, etc. Equation (5) thus dictates that for a solitary inner state-meaning, its observables-representations are distributed exponentially.

Based on the assertion that the same representation can 'stand for' different inner states, we can now extend the obtained result to the case of a large number of inner states observed simultaneously (e.g. throughout the communication process) as follows:

$$f_0(z) = \int_0^{\infty} \varphi(\lambda)\lambda\, \mathrm{e}^{-\lambda z}\, d\lambda\,, \tag{7}$$

where $f_0(z)$ is the PDF of $z$, and $\varphi(\lambda)$ is the PDF of $\lambda$.

The latter generalization partially accounts for the fact that in the case of complex autonomous systems, a measured stochastic variable ($z$, in our case) reflecting a system's behavior is, as a rule, a sum of random variables, where each of the summands accounts for the system's behavior in a steady or stationary state with certain parameters of the system's internal regulatory mechanisms. $f_0(z)$ accordingly specifies the distribution function of the representation occurrence number for the inner states characterized by the existence of different microparameters shaping a single distribution of $\lambda$. However, since communication hinges on the coupling of systems of different nature (e.g. cognitive, social, and physical) and with different properties that should well be expected to cause different distributions of $\lambda$, equation (7) should be reformulated to a still more universal form:

$$f(z) = \sum_{i=1}^{M} c_i f_{0_i}(z), \tag{8}$$

where $M$ is the number of distributions of $\lambda$, and $c_i$ gives the probability to observe the '$i$-th type' states (that is, in the context of our study, identical to observing the '$i$-th type' systems) in communication. It is important to note that in the statistical sense, $n$ the depth of coupling (see Definition IV) determines and theoretically limits $M$.

So far, we did not make any assumption about $\varphi(\lambda)$ the PDF of the composite parameter $\lambda$, and our result – the distribution equation (8) – is fairly general but is hard to validate or apply in practice because it yields no specific functional form. It is clear from equation (6) that as long as $\theta$ and $\rho$ are constant, the distribution of $\lambda$ depends on that of the representation time $\varepsilon$. The latter is a stochastic variable determined by, at least in the case of human communication, the durations of higher nervous activities, the exact distribution function of which is not yet known. There is increasing evidence from neurophysiology and psychophysics, however, that a Gamma distribution provides a reasonably good approximation (see Luce 1986). By substituting the Gamma PDF into equation (7), and after specializing and integrating equation (8), we finally obtain:

$$f(z) = \sum_{i=1}^{M} c_i \frac{\nu_i b_i^{\nu_i}}{(z+b_i)^{\nu_i+1}}, \tag{9}$$

where $b_i$ and $\nu_i$ are distribution parameters (it can be shown that $\frac{\nu_i}{b_i} \propto \frac{1}{\theta_i \rho_i} \mathrm{E}[\varepsilon_i]$, $\mathrm{E}[\varepsilon]$ denotes the mean of $\varepsilon$ ), and $i$=1,…,$M$.

To summarize: the probability density function of *z* the representation occurrence number in communication is specified by the sum (9) that is a result of the process of diffusion of *a*) *observationally* – inner states over their representations, or *b*) *internally* (e.g. *interpretationally*) – representations over their inner states, with parameters dependent on higher nervous activities ($\varepsilon$), behavioral characteristics ($\theta$), and efficiency of representation ($\rho$).

*Experiment*

To explore our theoretical findings and validate the resultant statistical model, we have conducted an experiment and compared probabilities of a word predicted by the model (9) with word actual occurrence frequencies in different text- and hypertext-based communications. Two samples of text (corpora) and two – of hypertext data have been analyzed. A set of short science fiction stories written by different authors in English were randomly selected and downloaded from `http://www.planetmag.com` to form an English text collection; a Russian text collection was analogously created from texts posted at `http://www.lib.ru/INOFANT`. Two collections of hypertext included connected (within the site) Web-pages randomly selected and downloaded from `http://www.cnn.com` and from `http://www.cbsnews.com`. All the texts were raw texts (English and Russian, respectively), where we ignored the punctuation and turned all word-forms into lower-case. Different word-forms were left unchanged and treated as different representations. All the hypertexts were stripped of multimedia and technical contents (including commercial ads), so that only English texts remained, which were then preprocessed in the same way as the two text collections.

We calculated the occurrence number of words in the prepared texts and hypertexts that made up four data samples. These data samples thus contain results of the measurement of the occurrence frequency of a representation (i.e. word) performed at different times in two hypermedia- and two text- based communications.

Maximum likelihood estimators (see Walpole and Myers 1993) were used to evaluate the parameters of the theoretical model (9), which was appropriately adjusted to deal with discrete data. The coefficients $c_i$ as well as the number of summands $M$ (that can be interpreted as the empirically discerned depth of coupling) were sought through a pareto-optimization procedure (Pareto-optimal means that it is not possible to make one parameter better fit without making other parameters worse fit) by minimizing $\chi^2$ the chi-square goodness-of-fit statistic. Figure 5 shows results of the experiment in the graphical form along with characteristics of the data samples and calculated parameters of the model.

It has been found that $M = 2$ provides a statistically sound fit: correlation coefficient $r \sim 0.99$ (i.e. the plots are clearly linear lines), together with the slope $\sim 1.0$ in all four plots, permits us to claim that the real data and the synthetic (i.e. calculated based on the model) data are from the same distribution. It is interesting to note that the calculated values of the $c_i$ coefficients are in a good agreement with the fact that natural language words can be classified into 'service/function' and 'content' words, which roughly make up 40-45% and 55-60% of the lexicon, respectively (Naranan and Balasubrahmanyan 1992). We could speculate that service/function words are indeed 'products,' in the statistical sense, of the social system, whereas content words are 'products' of psychic systems.

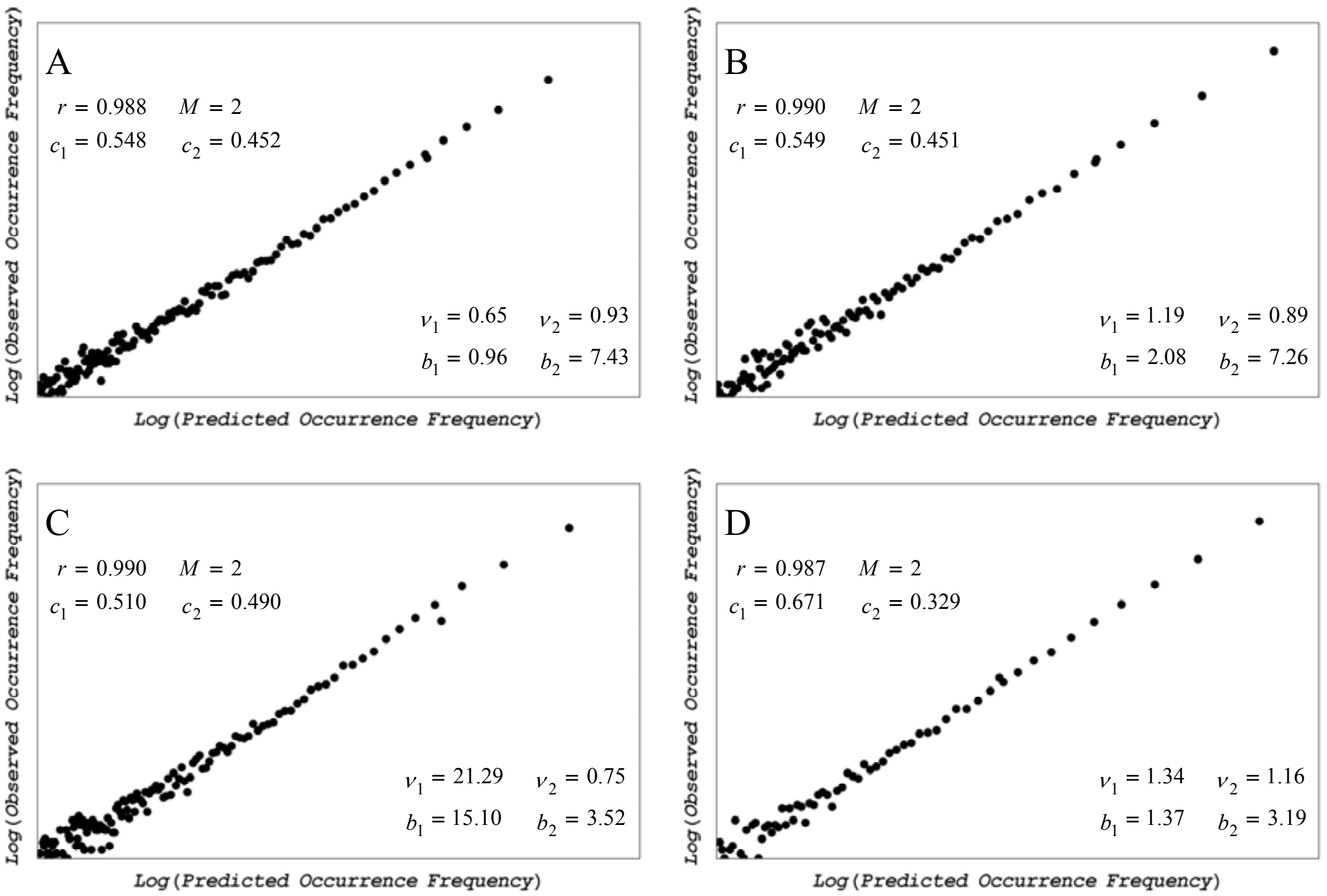

**Figure 5:** Histograms of the predicted occurrence frequency versus the observed occurrence frequency plotted on a log-log scale: A – CNN hypertexts (7.3Mb, vocabulary size ~ 1.0mln. words), B – CBS hypertexts (11.4Mb, 1.8mln. words), C – English texts (6.5Mb, 1.1mln. words), and D – Russian texts (3.1Mb, 0.45mln. words).

The obtained results imply that from the standpoint of mathematically modeling the communication process, there is no difference between Peirce's (triadic) and de Saussure's (dyadic) conceptualizations of the sign relationship for the considered communications, because 2 is the number of statistically recognizable components forming the representation dynamics of communication and, hence, the dyadic – two-component – modeling is already sufficient in the statistical sense. Besides, given the very general character of the theoretical model derivation and the plainness of the preprocessing procedure, the experimental results permit us to claim that fluctuations of representation frequencies follow the law written with equation (9) not only in 'natural' letter-by-letter/word-by-word, but any other *representation-based* communication.

**A Little Related Work and Discussion**

Although in jeopardy of being accused of reductionism, this paper has demonstrated, however unequivocally, that quantification and mathematical discourse can contribute to semiotics studies. On the other hand, it also demonstrated that the 'quantification' (though, given the relativistic character of the formalism, not 'mechanization') of the semiotic discourse does open new horizons in information system design, communication, and complex system theory. As a truly interdisciplinary research, the presented study covers a number of domains, and an adequate survey of related work in semiotics, cognitive science, sociology, linguistics, physics, computer science, etc. would quickly exceed the frames of one article. We thus selected for outline here only the works most important and/or influential to our theory. This should add to the discussion carried out throughout the previous sections.

In view of the two questions risen in Introduction – whether the realm of the 'mental' is that of the natural sciences, and what would be a foundation for a general communication theory, it was chiefly Andersen's 'Dynamic Semiotics' (Andersen 2002) that motivated us to look for a solution by compromising, refining, and (re-) defining semiotic concepts from positions of sociology and system theory. In pragmatically pursuing the development of theoretical principles for the design and analysis of hypermedia systems, we utilized the system-theoretic view of a complex system's representational closure first devised by Francis Heylighen (Heylighen 1990). Our quantum-physics-inspired axiomatic basis further develops – mathematically particularizes but expands to the general case of system theory – Gibson's concept of *affordances* (Gibson 1979), which are behavioral patterns of physical or biological significance, and Stamper's *norms* (Stamper 1996), which are the social counterpart of affordances (i.e. socially created patterns of behavior): these two are understood as observables of autonomous system inner states. Furthermore, our vision of a social system is rather close to Stamper's definition of an organization – a group of people who not merely share rules of language, customs, and habits, but also actively (due to the coupling) participate in the social construction of these rules. Our definition of communication rests on the qualitative description of the process given by di Paolo in his original work on the evolution of languages (di Paolo 1998), and it conceptually resembles Niklas Luhmann's declaration that communication is not transmission but the recursive 'use' of signs to generate signs, as information is an inner (psychic or conscious) state change, which does not exist externally and therefore cannot be transmitted (Luhmann 1995). We however do not adhere Luhmann's isolationistic vision of communication itself as a self-organizing system, but rather view it as a (by-) product of the coupling of such systems. Our theorization is therefore closer to Maturana and Valera's discussion of communication and '*languaging*' (Maturana and Varela 1980) as coordination of coordination (through the coupling of different type systems, in our case), and to the mainstream of Semiotics of Systems (Rocha 2000; Lemke 2000) and Second Order Cybernetics (Heylighen 2003).

Philosophically speaking, this study, although it did not completely deny the canonical hypothesis of 'ever unbroken' semiotic triad (as stated in Peirce 1998), it came to a principally new level of presentation of semiosis processes, specifically – the semiosis of communication, by shifting the focus to the (a)synchronicity in the developments of objects, interpretants, and signs. We demonstrated that while communication as coordinated behavior builds on the existence, at least potential, of completed semiotic triads, the dyadic modeling in terms of signifiers and signifieds (i.e. without explicitly referring to the environment) should, in many practical cases, suffice to analyze miscellaneous communication phenomena. In this way, our work somewhat opposes the recently popular (but rarely empirically supported) opinion in computer science that the triadic analysis scheme has obvious advantages when describing information transmission processes (e.g. see Condon 2000).

Now turning to the quantification of the semiotic account, we must mention the work by Ed Stephan, a sociologist, who convincingly demonstrated the power of the methods and tools of statistical physics in social studies (Stephan 2004). By adopting the metaphor of diffusion, we derived a formal (and hence computational), enclosed, and self-sufficient dynamic model of (representation-based) communication, which thereby does not depend directly on any other communication model, but can itself be specialized by taking on results obtained in other disciplines, such as physiology and linguistics. This model, along with the formulated axioms, creates the basis for the development of a general (system-theoretic semiotic) communication theory, which would ultimately replace the classical conveyor-tube framework.

Our experiments are straightforward but have a number of advantages over previous work. Analysis of communicability of hypermedia systems (such as the entire World-Wide Web or distributed multimedia databases) has been a hot research topic in theoretical computer science in the past decade (de Souza 1993; Goguen 1999; Condon 2000), yet so far with surprisingly few consequences for information system design: while the sophistication of design models has been growing, their practical value remains questionable, owing to the lack of formalization of the models' institutional concepts. The simple set-theoretic measure of behavioral coordination proposed in this paper proved a meaningful and, we believe, useful characteristic of the communication process (notably that similar but essentially static measures, such as Jaccard Coefficient, are long known in taxonomic studies). Statistical models of the representational dynamics were, on the other hand, abundant in quantitative linguistics over the past 50 years (for a bibliography, see Glottometrics 2002). Again surprisingly however, none of them is general enough to describe hypermedia-based communication, and none of them is statistically sound to accord with raw (i.e. without artful preprocessing but 'as is') empirical data. The model written with formulas (3-9) incorporates (in the formal sense, i.e. it can be specialized or reduced to) results of the earlier work in quantitative linguistics, such as by Mandelbrot (Mandelbrot 1960), and it demonstrates an excellent agreement between the theory and experiment. Besides, it methodologically provides a good starting point for a detailed and deeper analysis of the communication mechanisms (for instance, by explicitly assigning the initial conditions for the diffusion equation) and multimedia information systems (for instance, by varying the number of 'competing' media and/or their parameters). The latter could also help statistically explain or justify the rationale of the semiotic categorization of signs by their role (e.g. sign as 'symbol,' 'index,' or 'icon') as well as the 'reasonable' (i.e. statistically discernable) number of such sign clusters (or categories).

## Conclusion

Given the fact that theoretical analysis and empirical validation is fundamental to any model, whether conceptual or formal, it is simply astonishing that these two tools of scientific discovery are so often ignored in the contemporary studies of communication. In this paper, we pursued the ideas of *a*) correcting and expanding the modeling approaches of linguistics, which are otherwise inapplicable (more precisely, which should not but are widely applied), to the general case of hypermedia-based communication, and *b*) developing techniques for empirical validation of semiotic models, which are nowadays routinely used to 'explore' (in fact, to conjecture about) internal mechanisms of complex systems, yet on a purely speculative basis. This study thus offers two experimentally tested substantive contributions: the formal representation of communication as the mutually-orienting behavior of coupled autonomous systems, and the mathematical interpretation of the semiosis of communication, which together offer a concrete and parsimonious understanding of diverse communication phenomena. This understanding, however, still lacks many details, such as the explicit relationship between the state spaces and time dynamics of different type coupled systems. In terms of future work, the first obvious step would then be to clarify the definition of the *hierarchy* of the coupling systems and explore characteristic time scales on different levels of the hierarchy.

As a final (and probably 'far too late') remark, we would like to stress that as soon as computing and information systems have become a popular topic of semiotic research, further ignoring the formal aspects of semiotic models will make such research virtually useless. It is thus our hope that simply the presence of a handful of formulas in this paper will not alienate but rather encourage the reader to deeper into understanding semiosis and socio-semiotic phenomena and to sharpen the language surrounding the semiotic study of computer-mediated communication.

The presented work has been made within 'The Universal Design of the Digital City' project funded by the Japan Science and Technology Corporation.

**Caption list**

Figure 1**:** Information system and the transmission of messages.

Figure 2: Semiosis of hypermedia-based communication.

Figure 3: Coupling of autonomous systems (the case of $n = 2$).

Figure 4: Hypermedia-based communication as behavioral coordination.

Figure 5: Histograms of the predicted occurrence frequency versus the observed occurrence frequency plotted on a log-log scale: A – CNN hypertexts (7.3Mb, vocabulary size ~ 1.0mln. words), B – CBS hypertexts (11.4Mb, 1.8mln. words), C – English texts (6.5Mb, 1.1mln. words), and D – Russian texts (3.1Mb, 0.45mln. words).

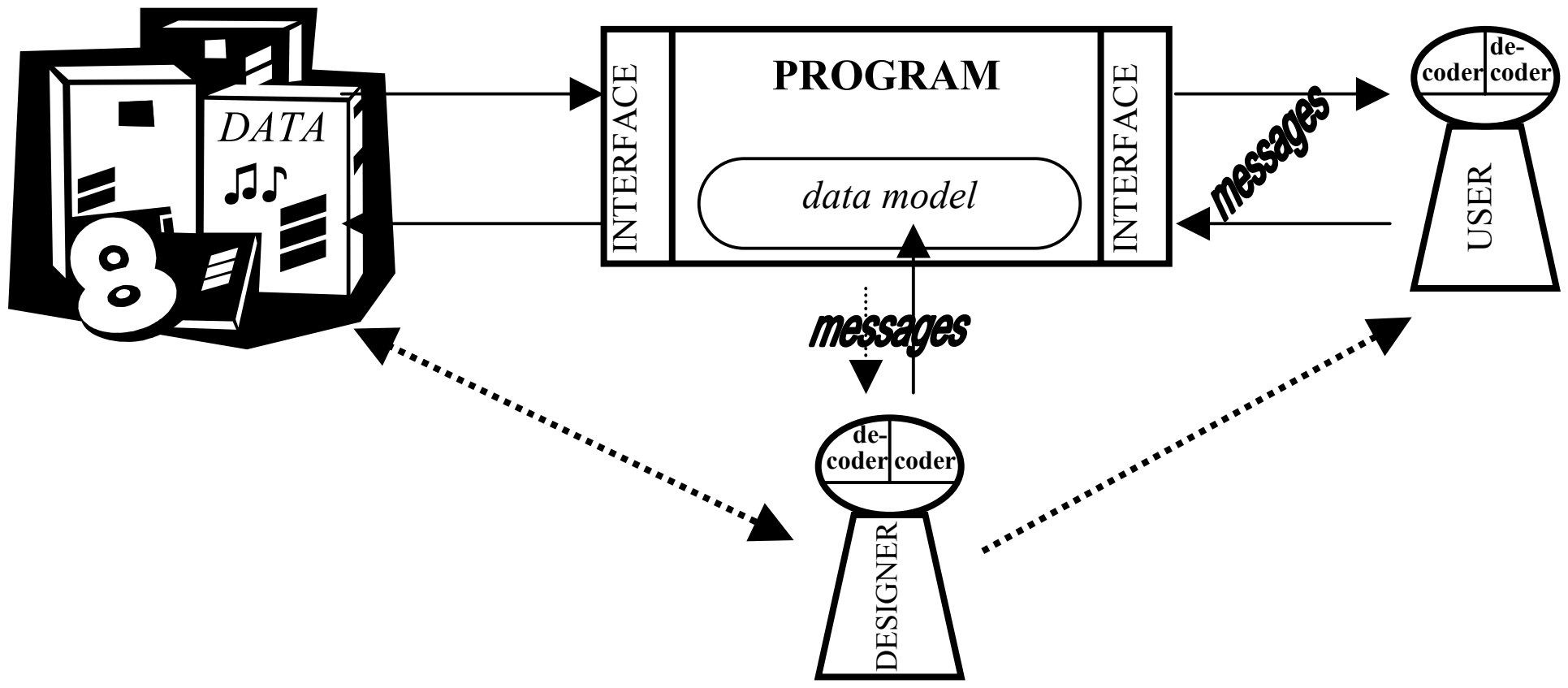
DATA
PROGRAM
INTERFACE
data model
INTERFACE
messages
coder
de-coder
USER
messages
de-coder
coder
DESIGNER

**Figure 1.**

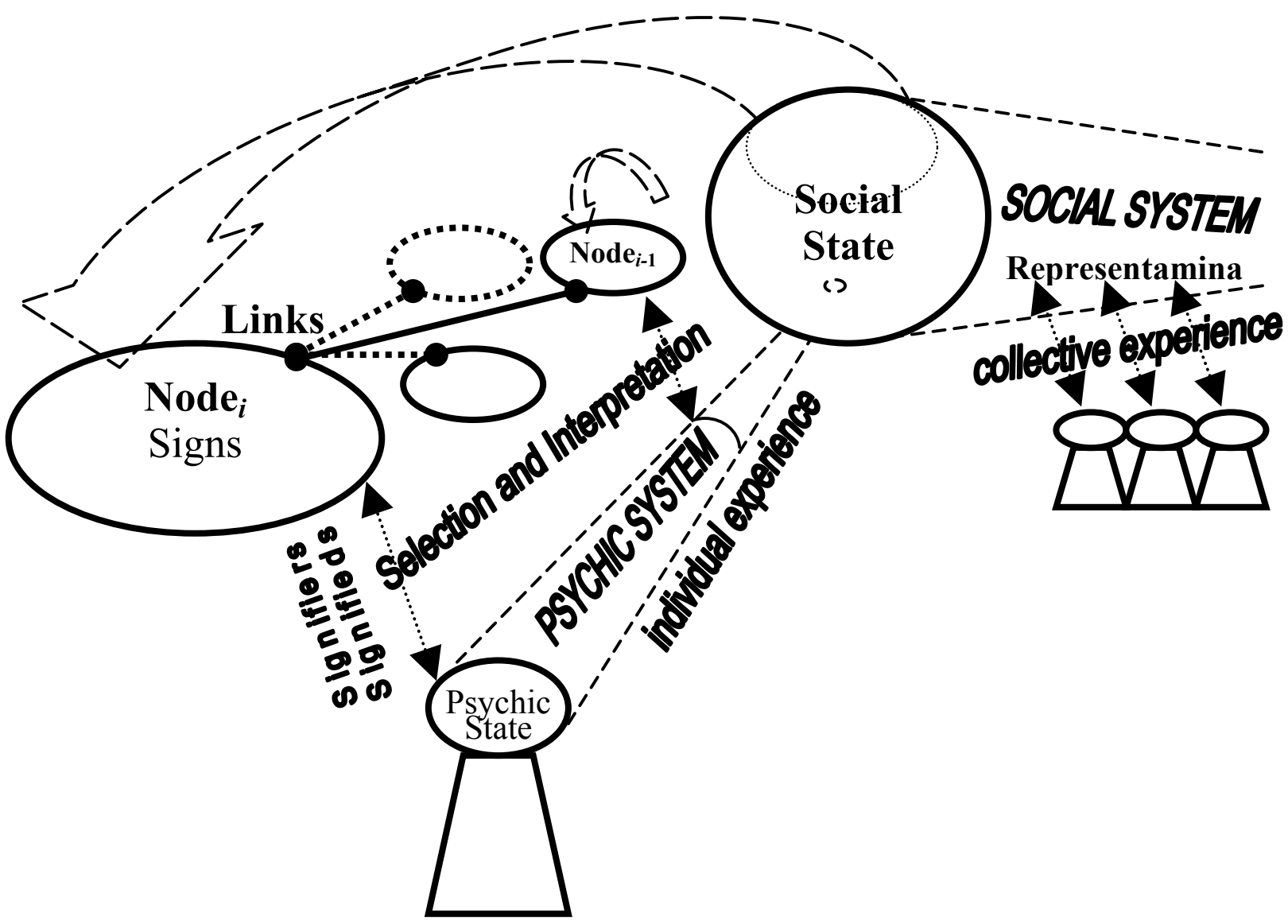

Social
State
SOCIAL SYSTEM
Representamina
collective experience
Node$_{i-1}$
Links
Node$_i$
Signs
Selection and Interpretation
PSYCHIC SYSTEM
individual experience
signifiers
signifieds
Psychic
State

**Figure 2.**

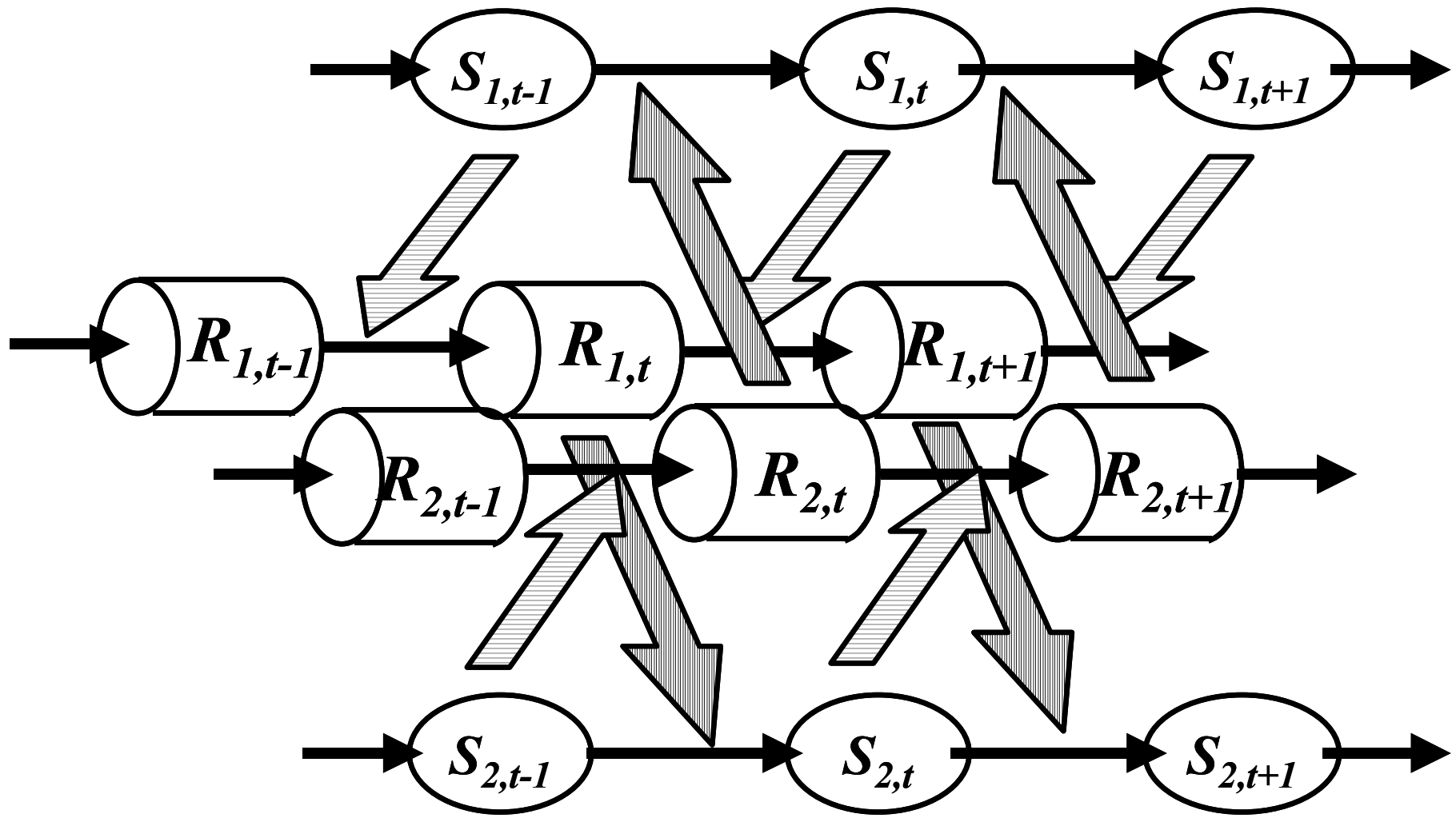

**Figure 3.**

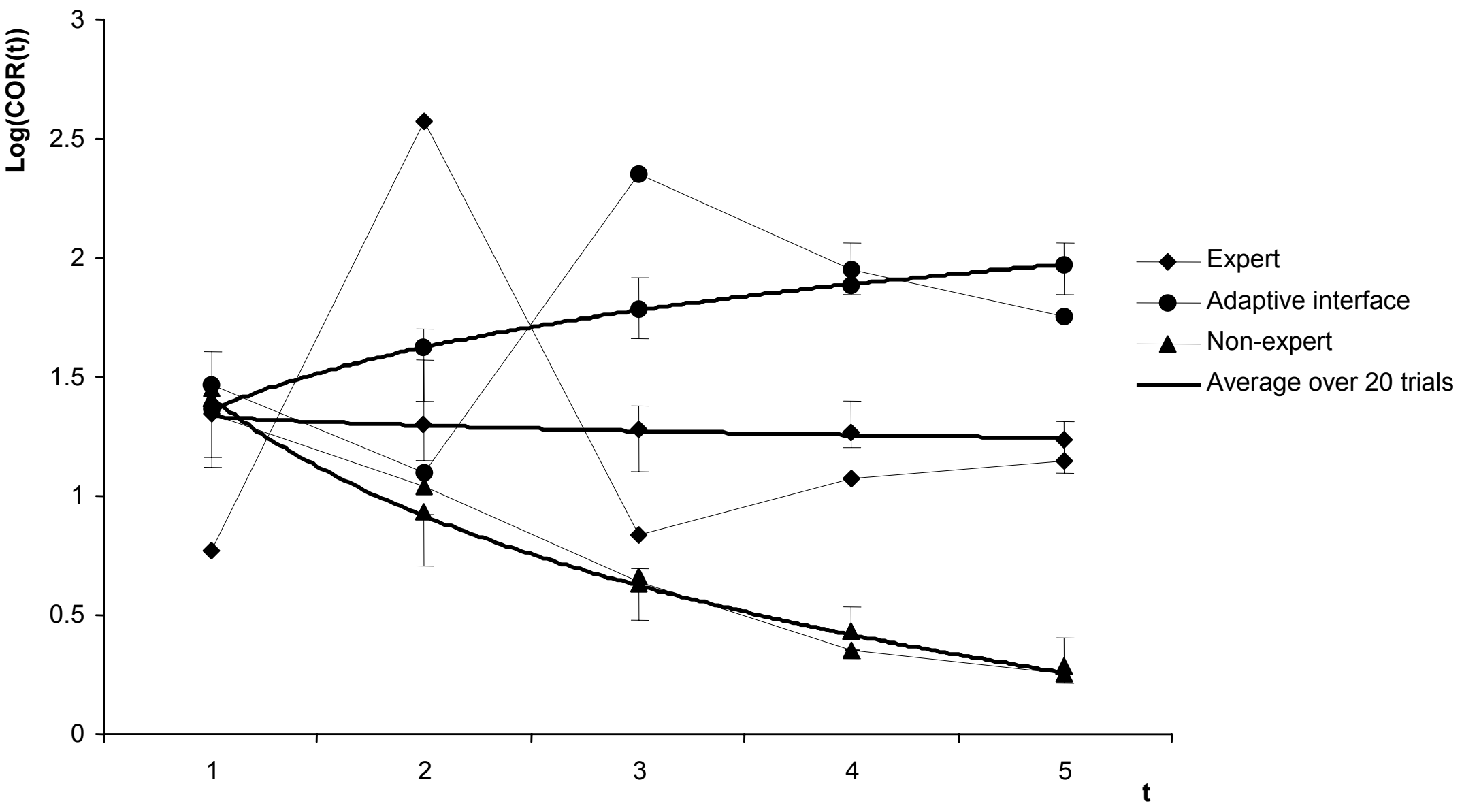

**Figure 4.**

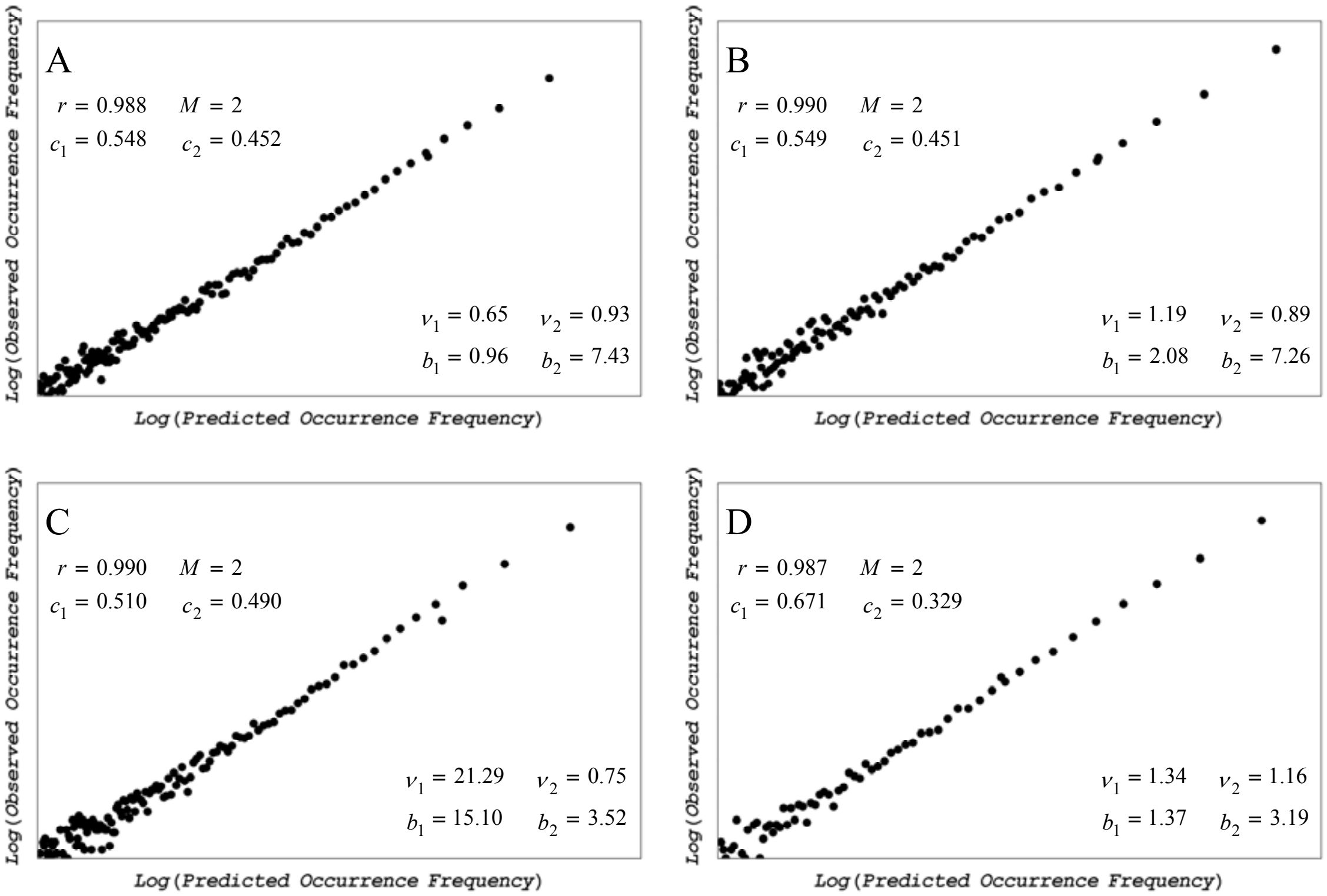
A
r = 0.988 M = 2
c1 = 0.548 c2 = 0.452
ν1 = 0.65 ν2 = 0.93
b1 = 0.96 b2 = 7.43
B
r = 0.990 M = 2
c1 = 0.549 c2 = 0.451
ν1 = 1.19 ν2 = 0.89
b1 = 2.08 b2 = 7.26
C
r = 0.990 M = 2
c1 = 0.510 c2 = 0.490
ν1 = 21.29 ν2 = 0.75
b1 = 15.10 b2 = 3.52
D
r = 0.987 M = 2
c1 = 0.671 c2 = 0.329
ν1 = 1.34 ν2 = 1.16
b1 = 1.37 b2 = 3.19
Log(Observed Occurrence Frequency)
Log(Predicted Occurrence Frequency)

**Figure 5.**